\documentclass[sigconf]{acmart}
\usepackage{color}
\usepackage{xcolor, colortbl}
\usepackage{multirow}
\usepackage[utf8]{inputenc}
\usepackage[english]{babel}

\definecolor{mygray}{gray}{.8}

\usepackage[linesnumbered,ruled,vlined]{algorithm2e}
\usepackage{caption}
\usepackage{subcaption, amsmath}
\usepackage{paralist}
\usepackage{enumitem}
\usepackage{bbold}

\makeatletter
\newcommand{\removelatexerror}{\let\@latex@error\@gobble}
\makeatother
\AtBeginDocument{%
  \providecommand\BibTeX{{%
    \normalfont B\kern-0.5em{\scshape i\kern-0.25em b}\kern-0.8em\TeX}}}

\acmConference[SSDBM '20]{SSDBM '20: Int. Conf. on Scientific and Statistical Data Management}{June 03--05, 2020}{Amsterdam, Netherlands}
\acmBooktitle{SSDBM '20: Int. Conf. on Scientific and Statistical Data Management,
  June 03--05, 2020, Amsterdam, Netherlands}
\acmPrice{15.00}
\acmISBN{978-1-4503-XXXX-X/18/06}

\acmSubmissionID{xxx}

\begin{document}

\title{An Algebraic Approach for High-level Text Analytics}

\author{Xiuwen Zheng}
 \email{xiz675@eng.ucsd.edu}
 \affiliation{%
  \institution{University of California San Diego}
  \streetaddress{San Diego Supercomputer Center}
  \city{La Jolla}
  \state{Californa, USA}
  \postcode{92130}
}
 \author{Amarnath Gupta}
\email{a1gupta@ucsd.edu}
%\orcid{0000-0003-0897-120X}
\affiliation{%
  \institution{University of California San Diego}
  \streetaddress{San Diego Supercomputer Center}
  \city{La Jolla}
  \state{Californa, USA}
  \postcode{92130}
}

\renewcommand{\shortauthors}{X. Zheng}

\begin{abstract}
Text analytical tasks like word embedding, phrase mining and topic modeling, are placing  increasing demands as well as challenges to existing database management systems. 
%Further, any analytical task typically involves workflows that include standard relational operations together with text-specific operations. 
In this paper, we provide a novel algebraic approach based on associative arrays. Our data model and algebra can bring together relational operators and text operators, which enables interesting  optimization opportunities for hybrid data sources that have both relational and textual data. We demonstrate its expressive power in text analytics using several real-world tasks. 
\end{abstract}


\maketitle

\section{Introduction}
\label{sec:intro}
A significant part of today's analytical tasks involve text operations. A data scientist who has to manipulate and analyze text data today typically uses a set of text analysis software libraries (e.g., NLTK, Stanford CoreNLP, GenSim) for tasks like word embedding, phrase extraction, named entity recognition and topic modeling. In addition, most DBMS systems today have built-in support for full-text search. PostgreSQL, for instance, admits a text vector (called tsvector) that extracts and creates term and positional indices to enable efficient queries (called tsquery). Yet, some common and seemingly simple text analysis tasks cannot be performed simply within the boundaries of a single information system.

\noindent \textbf{Example 1.} Consider a relational table 
$R$(\underline{newsID}, date, newspaper, title, content) 
where $title$ and $content$ are text-valued attributes, and two sets $L_o, L_p$ that represent a collection of organization names and person names respectively. Now, consider the following analysis:
\begin{itemize}[leftmargin=*]
    \item $N1 =$ Select a subset of news articles from date $d_1$ through $d_2$
    \item $N2 =$ Identify all news articles in $N1$ that have at least $c_1$ organization names from $L_o$ \underline{and} $c_2$ persons from $L_p$
    \item $T1 =$ Create a document-term matrix on $N2.text$
    \item $T2 =$ Remove rows and columns of the matrix if either of their row or column marginal sums is below $\theta_1$ and $\theta_2$ respectively.
    \item $M = $ Compute a topic model using $T2$
\end{itemize}
The intention of the analysis is to find the topic distribution of those news items that cover, for example, any two members of the senate (list $L_p$) and any one government organizations (list $L_o$). The analysis itself is straightforward and can be performed with a combination of SQL queries and Python scripts. 

Our goal in this short paper is to present the idea that a novel \textit{relation-flanked associative array data model} has the potential of serving as the underlying framework for the management and analysis of text-centric data. We develop the theoretical elements of the model and illustrate its utility through examples.

\section{The Data Model}
\subsection{Text Associative Arrays}~\label{subsec:associative_array}
A number of current data systems, typically in the domain of polystore data systems, use associative arrays \cite{jananthan2017polystore,kepner2020ai} or its variants like associative tables \cite{barcelo2019expressiveness} and tensor data model \cite{leclercq2019polystore}. Many of these data models are used to support analytical (e.g., machine learning) tasks. In our setting, we specialize the essential associative model for text analytics. For our level of abstraction, our model reuses relational operations for all metadata of the associative arrays. While it has been shown \cite{barcelo2019expressiveness} that associative arrays can express relational operations, we believe that using relational abstraction along with our text-centric algebraic operations makes the system easier to program and interpret. At a more basic level, since most text processing operations include sorting (e.g., by TF-IDF scores), our model is based on partially ordered semirings.
\begin{definition}[Semiring]~\label{def:semiring}
    A semiring is a set $R$ with two binary operations addition $\oplus$ and multiplication $\odot$, such that, 1) $\oplus$ is associative and commutative and has an identity element $0\in R$; 2) $\odot$ is associative with an identity element $1\in R$; 3) $\odot$ distributes over $\oplus$; and 4) $\odot$ by 0 annihilates $R$.
\end{definition}

\begin{definition}[Partially-Ordered Semiring]~\cite{golan2013semirings}\label{def:pos}
A semiring $R$ is partially ordered if and only if there exists a partial order
relation $\leq$ on $R$ satisfying the following conditions for all $a, b \in R$:
    \begin{itemize}
        \item If $a\leq b$, then $a\oplus c \leq b\oplus c$;
        \item If $a\leq b$ and $0\leq c$, then $a\odot c\leq b\odot c$ and $c\odot a\leq c\odot b$.
    \end{itemize}
\end{definition}

\begin{definition}[Text Associative Array]~\label{def:ass_array}
    The Text Associative Array (TAA) $\mathbf{A}$ is defined as a mapping:
    \begin{displaymath}
        \mathbf{A}: K_1 \times K_2 \to R
    \end{displaymath}
    where $K_1$ and $K_2$ are two key sets (named row key set and column key set respectively), and $R$ is a partially-ordered semiring (Definition~\ref{def:pos}). We call $K_1\times K_2$ ``the dimension of  $\mathbf{A}$'', and denote $\mathbf{A}.K_1$, $\mathbf{A}.K_2$ and $\mathbf{A}.K$ as the row key set, column key sets, and set of key pairs of $\mathbf{A}$, respectively.
\end{definition}

Next, we define the basic operations on text associative arrays, to be used by our primary text operations (Sec.~\ref{subsec:text_operations}).

\begin{definition}[Addition]~\label{def:add}
    Given two TAAs $\mathbf{A},\mathbf{B}: K_1\times K_2\to R$, the addition operation $\mathbf{C}=(\mathbf{A}\oplus\mathbf{B}):K_1\times K_2\to R$ is defined as,
    \begin{displaymath}
        \mathbf{C}(k_1,k_2) = (\mathbf{A}\oplus\mathbf{B})(k_1,k_2) = \mathbf{A}(k_1,k_2)\oplus\mathbf{B}(k_1,k_2).
    \end{displaymath}
    Define $\mathbb{0}_{K_1,K_2}$ as a TAA where $\mathbb{0}_{K_1,K_2}(k_1,k_2)=0$ for $\forall k_1\in K_1, k_2\in K_2$. $\mathbb{0}_{K_1,K_2}$ serves as an identity for addition operation on key set $K_1\times K_2$.
\end{definition}

\begin{definition}[Hadamard Product]~\label{def:dot_mul}
    Given two TAAs $\mathbf{A},\mathbf{B}: K_1\times K_2\to R$, the Hadamard product operation $\mathbf{C}=(\mathbf{A}\odot\mathbf{B}):K_1\times K_2\to R$ is defined as,
    \begin{displaymath}
        \mathbf{C}(k_1,k_2) = (\mathbf{A}\odot\mathbf{B})(k_1,k_2) = \mathbf{A}(k_1,k_2)\odot \mathbf{B}(k_1,k_2).
    \end{displaymath}
    Define $\mathbb{1}_{K_1,K_2}$ as a TAA where $\mathbb{1}_{K_1,K_2}(k_1,k_2)=1$ for $\forall k_1\in K_1, k_2\in K_2$. $\mathbb{1}_{K_1,K_2}$ serves as an identity for Hadamard product on key set $K_1\times K_2$.
\end{definition}

\begin{definition}[Array Multiplication]~\label{def:array_mul}
    Given two TAAs $\mathbf{A}: K_1\times K_2\to R$ and $\mathbf{B}: K_2\times K_3\to R$, the array multiplication operation $\mathbf{C}=(\mathbf{A}\otimes\mathbf{B}):K_1\times K_3\to R$ is defined as,
    \begin{displaymath}
        \mathbf{C}(k_1,k_3) = (\mathbf{A}\otimes\mathbf{B})(k_1,k_3) = \bigoplus_{k_2\in K_2}\mathbf{A}(k_1,k_2)\odot\mathbf{B}(k_2,k_3).
    \end{displaymath}
\end{definition}

\begin{definition}[Array Identity]~\label{def:array_identity}
Given two key sets $K_1$ and $K_2$, and a partial function $f: K_1 \hookrightarrow K_2$, the array identity  $\mathbb{E}_{K_1,K_2, f}:K_1\times K_2\to R$ is defined as a TAA  such that
    $$
        \mathbb{E}_{K_1,K_2, f}(k_1,k_2) = 
        \begin{cases}
            1, & \text{if }k_1\in \text{dom } f \text{ and }k_2=f(k_1);\\
            0, & \text{otherwise}.
        \end{cases}
    $$
    Specifically, if $\text{dom }f = K_1\cap K_2$ and $f(k_1) = k_1$ for $\forall k_1 \in K_1$, $\mathbb{E}_{K_1,K_2,f}$ is abbreviated to $\mathbb{E}_{K_1,K_2}$.
\end{definition}
In general, $\mathbb{E}_{K_1,K_2, f}(k_1,k_2)$ is not an identity for general array multiplication. However, $\mathbb{E}_{K,K}$ is an identity element for array multiplication on associative arrays $K\times K\to R$.

\begin{definition}[Kronecker Product]~\label{def:kronecker_product}
    Given two TAAs $\mathbf{A}:K_1\times K_2\to R$ and $\mathbf{B}:K_3\times K_4\to R$, their Kronecker product $\mathbf{C}=\mathbf{A}\circledast\mathbf{B}:(K_1\times K_3)\times(K_2\times K_4)$ is defined by
    $$
        \mathbf{C}((k_1,k_3),(k_2,k_4)) = \mathbf{A}(k_1,k_2)\odot\mathbf{B}(k_3,k_4).
    $$
\end{definition}

\begin{definition}[Transpose]~\label{def:transpose}
    Given a TAA $\mathbf{A}:K_1\times K_2\to R$, its transpose, denoted by $\mathbf{A}^{\mathsf{T}}$, is defined by $\mathbf{A}^{\mathsf{T}}:K_2\times K_1\to R$ where $\mathbf{A}^{\mathsf{T}}(k_2,k_1)=\mathbf{A}(k_1,k_2)$ for $k_1\in K_1$ and $k_2\in K_2$.
\end{definition}

\subsection{Text Operations}~\label{subsec:text_operations}
We can express a number of fundamental text operations using the proposed TAA algebra. We first define three basic TAAs specifically for text analytics, then a series of text  operations will be defined  on general TAA or  these basic structures.

\begin{definition}[Document-Term Matrix]~\label{def:dom}
    Given a text corpus, a document term matrix is defined as a TAA $\mathbf{M}: D\times T\to R$ where $D$ and $T$ are the document set and term set of a text corpus. 
\end{definition}
The term set  in the document-term matrix can be the vocabulary or the bigram of the corpus, or an application-specific user-defined set of interesting terms. The matrix value  $\mathbf{M}(d, t)$ can also take different semantics, in one application it can be the occurrence of $t$ in document $d$, while in another application, it can be the term frequency-inverse document frequency (tf-idf). Typically, elements of $D$ and $T$ will have additional relational metadata. A document may have a date and a term may have an annotation like a part-of-speech (POS) tag.

\begin{definition}[Term-Index Matrix]~\label{def:tim}
   Given a document $d$, the term index matrix is defined as a TAA, $\mathbf{N}: T_d\times I\to \{0,1\}$ where $T_d=\{d\}\times T$ is the set of terms in document $d$ and $I=\{1,\cdots,I_d\}$ is the index set ($I_d$ is the size of $d$). Specifically, for $(d, t)\in T_d$ and $i\in I$,
    $$
        \mathbf{N}((d,t), i) =
        \begin{cases}
            1, &\text{if }i\text{-th word of document }d\text{ is }t;\\
            0, &\text{otherwise}.
        \end{cases}
    $$
\end{definition}

\noindent \textbf{Example 2. }
For a document $d$ = ``Today is a sunny day'', let its term index matrix be $\mathbf{N}: (\{d\}\times T)\times I \to \{0, 1\}$, then we have $T = \{\text{``today'',        is'',  ``a'', ``sunny'', ``day''}\}$, $I = \{1, 2, 3, 4, 5\}$. $\mathbf{N}(\text{``today''}, 1) = 1, \mathbf{N}(\text{``is''}, 2) = 1, \mathbf{N}(\text{``a''}, 3)=1, \mathbf{N}(\text{``sunny''}, 4) = 1,\mathbf{N}(\text{``day''}, 5) = 1$, and for other $(t, i)$ pairs where $(t, i)\in T\times I$, we have $\mathbf{N}(t, i) = 0$.  

\begin{definition}[Term Vector]~\label{def:tv}
    There are two types of term vectors. 1) Given a set of terms $T$ of a document $d$, the term vector is defined as a TAA $\mathbf{V}:\{d\}\times T \to R$. 2) Given a set of terms $T$ for a collection of documents $D$, $\mathbf{V}:\{1\}\times T \to R$ is  a term vector for the corpus $D$.
\end{definition}
The term vector represents some attribute of  terms in the scope of one document or a corpus. For example, for a document $d$, the value of the  term vector $\mathbf{V}:\{d\}\times T$ can be the occurrence of each term in this document.  For a corpus $D$, the value of its term vector $\mathbf{V}:\{1\}\times T$ can be idf value for each term in the whole corpus, and the value is not specific to a single document.  

Based on these structures, we can define our unit text operators as follows. Some operators are defined for general TAAs, while some are defined for a specific type of TAAs. 

\begin{definition}[Extraction]~\label{def:extract}
    Given a TAA  $\mathbf{A}:K_1\times K_2\to R$ and two projection sets $K_1'\subseteq K_1$, $K_2'\subseteq K_2$, we define the extraction operation as 
    $$
        \Pi_{K_1',K_2'}(\mathbf{A}) = \mathbb{E}_{K_1',K_1}\otimes\mathbf{A}\otimes\mathbb{E}_{K_2',K_2}^\mathsf{T}.
    $$
    Let $\mathbf{B} = \Pi_{K_1',K_2'}(\mathbf{A})$, we have $B(k_1, k_2) = A(k_1, k_2), \text{ for } \forall (k_1, k_2) \in K_1' \times K_2'$.
\end{definition}
When only extracting row keys, the operation can be expressed as $\Pi_{K_1', :}$ and when extracting column keys, it is expressed as $\Pi_{:, K_2'}$.

\begin{definition}[Rename]~\label{def:rename}
    Given a TAA $\mathbf{A}:K_1\times K_2\to R$, suppose $K_2'$ is another column key set and there exists a bijection $f: K_2\to K_2'$. The column rename operation is defined as 
    $$
    \rho_{K_1,K_2\to K_2', f}(\mathbf{A})=\mathbf{A}\otimes \mathbb{E}_{K_2,K_2',f}.
    $$
    Similarly, given another row key set $K_1'$ and a bijection $f:K_1\to K_1'$, the row rename operation is defined as
    $$
    \rho_{K_1\to K_1',K_2, f}(\mathbf{A})=\mathbb{E}_{K_1',K_1,f^{-1}}\otimes\mathbf{A}.
    $$
    The subscript $f$ can be omitted if the bijection is clear, e.g., $|\text{dom }f| = 1$. In addition, the row rename operation and column rename operation can be combined together as $\rho_{K_1\to K_1',K_2\to K_2'}(\mathbf{A})$. Our rename operator is more general than the rename operation of relational algebra since it supports both row key set and column key set renaming.
\end{definition}

\begin{definition}[Apply]~\label{def:apply}
Given a TAA $\mathbf{A}:K_1\times K_2\to R$ and a function $f:R\to R$, define the apply operator by
$\mathbf{Apply}_f(\mathbf{A}): K_1\times K_2 \to R$ where,
$$
    \mathbf{Apply}_f(\mathbf{A})(k_1, k_2) = f(\mathbf{A}(k_1, k_2)), \forall (k_1, k_2)\in K_1\times K_2.
$$
\end{definition}

\begin{definition}[Filter]~\label{def:filter}
    Given a TAA  $\mathbf{A}:K_1\times K_2\to R$ and an indicator function $f:R\to\{0,1\}$, define the filter operation on $\mathbf{A}$ as 
    $$
        \mathbf{B} = \mathbf{Filter}_f(\mathbf{A})=\sigma_f(\mathbf{A}) : K_{1f}, K_{2f} \to R,
    $$
    where $K_{1f}\times K_{2f} =\{(k_1,k_2)|(k_1,k_2)\in K_1\times K_2\text{ and }f(\mathbf{A}(k_1,k_2))=1\}$, and $\mathbf{B}(k_1, k_2) = \mathbf{A}(k_1, k_2)$.
\end{definition}

\begin{definition}[Sort]~\label{def:sort}
Given a TAA $\mathbf{A}:K_1\times K_2\to R$, for any $k \in K_1$, we extract a TAA $\mathbf{V}=\Pi_{\{k\},:}(\mathbf{A})$ of dimension $\{k\}\times K_2$. Since $R$ is a partially-ordered semiring (Definition~\ref{def:pos}), the value set $\{\mathbf{V}(k,x)|\forall x \in K_2 \} \subseteq R$ inherits the partial order from $R$,  which implies an order $\mathbf{V}(k,x_1)\leq \mathbf{V}(k,x_2)\leq \cdots \leq\mathbf{V}(k, x_{|K_2|})$. Define $\mathbf{Idx}(k, x_i)=i$, then the sort by column operation is defined as 
$$
    \mathbf{Sort}_2(\mathbf{A}):K_1\times K_2\to \{1,\cdots, |K_2|\},
$$ 
where $\mathbf{Sort}_2(\mathbf{A})(k,x)=\mathbf{Idx}(k, x)$.  Similarly, we have sort by row operation defined as
$$
    \mathbf{Sort}_1(\mathbf{A}):K_1\times K_2\to \{1,\cdots, |K_1|\}.
$$
When the column key dimension or row key dimension is 1 (e.g., for a term vector), $\mathbf{Sort}_1$ or $\mathbf{Sort}_2$ is abbreviated to $\mathbf{Sort}$.
\end{definition}

\begin{definition}[Merge]
    Given two TAAs $\mathbf{A}: K_{A1} \times K_{A2}$ and $\mathbf{B}: K_{B1}\times K_{B2}$, if $(K_{A1}\times K_{A2}) \cap (K_{B1}\times K_{B2}) = \emptyset$,
    then merge operation can be applied on them, and it is defined as, 
    $$
        \mathbf{C} = \mathbf{Merge}(\mathbf{A}, \mathbf{B}): K_1\times K_2 \to R
    $$
    where $K_1 =  K_{A1} \cup K_{B1}$ and  $K_2 = K_{A2}\cup K_{B2}$, and
    $$
    \begin{aligned}
        \mathbf{C}(k_1, k_2) = 
            \begin{cases}
                \mathbf{A}(k_1, k_2), & \text{if } (k_1, k_2) \in K_{A1} \times K_{A2};\\
                \mathbf{B}(k_1, k_2), & \text{if } (k_1, k_2) \in K_{B1} \times K_{B2};\\
                0, &\text{otherwise.}
            \end{cases}
    \end{aligned}
    $$
\end{definition}

\begin{definition}[Expand] ~\label{def:expand}
Given an elementwise binary operator $\mathbf{OP}$ on associative arrays, e.g., $\oplus$ and $\odot$, a term vector $\mathbf{V}: \{1\}\times T \to R$ and a document-term matrix $\mathbf{M}: D\times T \to R$, the expand operator is defined as 
$$
    \mathbf{Expand}_{\mathbf{OP}}(\mathbf{V}, \mathbf{M}) = \rho_{\{1\}\times D\to D, T\times\{1\}\to T}\left(\mathbf{V}\circledast\mathbb{1}_{D,\{1\}}\right)\,\,\mathbf{OP}\,\,\mathbf{M}.
$$
This operator implicitly expands the term vector $\mathbf{V}$ to generate another associative array $\mathbf{M'}: D\times T \to R$ where $\mathbf{M'}(d, t) = \mathbf{V}(1, t), \forall d\in D \text{ and } \forall t\in T$, and then applies $\mathbf{OP}$ on $\mathbf{M'}$ and $\mathbf{M}$. 
\end{definition}

Suppose that for a corpus $D$, there is a term vector $\mathbf{V}: \{1\}\times T\to R$ where $\mathbf{V}(1, t)$ is the mean occurrence of term $t$ in $D$ (i.e., $\frac{Count_t}{|D|}$ where $Count_t$ is the total occurrence of $t$ in $D$), and there is a document-term matrix $\mathbf{M}: D\times T$, then
$$
    \mathbf{Expand}_{\oplus}(\mathbf{Apply}_{f(x)=-x}(\mathbf{V}), \mathbf{M})
$$
will generate the difference of terms occurrences for each document from their average occurrences.  
% \begin{definition}[Concatenation]
% Concatenation operation is defined on two term vectors with the same row key, which is the special case for Merge operation. Let  $\mathbf{V_1}: \{d\}\times T_1 \to R, \mathbf{V_2}: \{d\} \times T_2 \to R$, define concatenation by, $\mathbf{V} = \mathbf{Concat}(\mathbf{V_1}, \mathbf{V_2}): \{d\}\times{T} \to R$ where $T = T_1 \cup T_2$.
% \end{definition}
\begin{definition}[Flatten]~\label{def:flatten}
    Given an associative array $\mathbf{A}:K_1\times K_2\to R$, the flatten operation is defined by $\mathbf{Flatten}(\mathbf{A}):\{1\}\times(K_1\times K_2)\to R$ where 
    $$
    \mathbf{Flatten}(\mathbf{A})(1,(k_1,k_2)) = \mathbf{A}(k_1,k_2) \text{ for } \forall (k_1,k_2)\in K_1\times K_2.
    $$
\end{definition}
\begin{definition}[Left Shift]~\label{def:shift}
    Given a term-index matrix $\mathbf{N}:(\{d\}\times T)\times I\to R$, and a non-negative integer $n$, define the left shift operator by 
    $\mathbf{LShift}_n(\mathbf{N}):(\{d\} \times T)\times I\to R$
    where 
    $$
    \begin{aligned}
        \mathbf{LShift}_n(\mathbf{N}) &=\mathbf{LShift}_1(\mathbf{LShift}_{n-1}(\mathbf{N}))\text{ and }\\
        \mathbf{LShift}_1(\mathbf{N})((d,t),i) &= 
        \begin{cases}
            \mathbf{N}((d, t),i+1), & \text{if }i<|T|;\\
            0, & \text{if }i=|T|;.
        \end{cases}
    \end{aligned}
    $$
\end{definition}
For a term-index matrix $\mathbf{N}$ of document $d$, $\mathbf{LShift}_1(\mathbf{N})$ generates another term-index matrix $\mathbf{N'}$ where $\mathbf{N'}((d, t), i) = 1$ when $t$ is the $(i+1)$-th word in $d$.  

\begin{definition}[Union]
    Suppose there are two term-index matrices with the same index set $I$, $\mathbf{N}_1: (\{d\}\times T)\times I \to R$ and $\mathbf{N}_2: (\{d\}\times T)\times I \to R$, the union operation on $\mathbf{N}_1$ and $\mathbf{N}_2$ is defined by
    $$
    \begin{aligned}
        \mathbf{Union}(\mathbf{N_1}, \mathbf{N_2})
        &= \rho_{(\{d\}\times T)\times(\{d\}\times T)\to \{d\}\times(T\times T), I\times I\to I}\\
        &\qquad\qquad\left( \Pi_{:, \{(i,i)|i\in I\}}(\mathbf{N}_1\circledast\mathbf{N}_2)\right).
    \end{aligned}
    $$
    Suppose $\mathbf{N}=\mathbf{Union}(\mathbf{N}_1,\mathbf{N}_2)$, then
    $$
        \begin{aligned}
            \mathbf{N}((d, (t_1, t_2)), i) = 
                \begin{cases}
                    1, & \text{if }\mathbf{N}_1((d, t_1), i) = 1 \text{ and }\mathbf{N}_2((d, t_2), i) = 1;\\
                    0, & \text{otherwise}.
                \end{cases}
        \end{aligned}
    $$
\end{definition}

The left shift and union operations can be composed to compute all bigrams of a document. Given a term-index matrix $\mathbf{N}$ of document $d$, let $\mathbf{N}' = \mathbf{Union}(\mathbf{N}, \mathbf{LShift}_1(\mathbf{N}))$, then $\mathbf{N}'((d, (t_1, t_2)), i) = 1$ when $(t_1, t_2)$ is the $i$-th bigram in document $d$.

\begin{definition} [Sum]
The sum operation takes  a TAA  $\mathbf{A}:K_1\times K_2 \to R$ and an integer  which can take the value of 0, 1 or 2 as inputs  and will have different semantics based on the  integer value:
\begin{align*}
 \mathbf{B}: \{1\} \times K_2 &=  \mathbf{Sum}_1(\mathbf{A}) \text{ where } B(1, k_2) = \bigoplus_{k_1\in K_1} \mathbf{A}(k_1, k_2);\\
  \mathbf{B}:{K_1} \times \{1\} &=  \mathbf{Sum}_2(\mathbf{A}) \text{ where } B(k_1, 1) = \bigoplus_{k_2\in K_2} \mathbf{A}(k_1, k_2).
\end{align*}
\end{definition}

% \subsection{Properties of Operators}

\section{Text Analytic Tasks}
\subsection{Constructing a Document Term Matrix}
As we state in Section~\ref{subsec:text_operations}, a document term matrix is a common representation model for a collection of documents where the terms can be a list of import terms or the whole vocabulary or bigrams. The entry of the matrix can be either the occurrence of each term or the tf-idf value.
%We provide two examples of building document term matrix and express these tasks using our text operators.

\noindent \textbf{Example 3. }
For document collection $C$, build a document term matrix where terms are all unigrams and bigrams in $C$, and the values should be the occurrence of each term in the whole corpus. 

Suppose there is a tokenization function called $\mathbf{Tokenize}$ that  takes a document $d$ as input and generates a term index matrix $\mathbf{N}: (\{d\}\times T)\times {I}$. 
The construction can be decomposed to two parts, the first part is to construct a Term  Vector for one single document $d$ containing all unigrams and bigrams together with their corresponding occurrences. Fig.~\ref{fig:bigram} shows the construction process.
%and note that we mark the dimension of associative array for the result of each step.

\begin{figure}[ht]
    \centering
    \begin{align*}
    &\mathbf{N} = \mathbf{Tokenize}(d)\quad :(\{d\}\times T)\times {I} & 1 \\
    & \mathbf{V_1} = \rho_{\{1\} \to \{d\}, \{d\}\times T \to T}(\mathbf{Sum}_2(\mathbf{N}))^\mathsf{T} \quad :\{d\}\times T & 2\\
    &\mathbf{T} = \mathbf{N} \otimes \mathbf{LShift}_1(\mathbf{N})^\mathsf{T} \quad :(\{d\}\times T)\times(\{d\}\times T)& 3\\
    &\mathbf{V_2} = \mathbf{Flatten}(T)\quad :\{1\}\times (\{d\}\times T)\times(\{d\}\times T))& 4\\
    &\mathbf{V_2} = \rho_{\{1\}\to\{d\},(\{d\}\times T)\times(\{d\}\times T)\to (T\times T)}(\mathbf{V}_2)\quad :\{d\}\times (T\times T)& 5\\
    &\mathbf{V_2} = \sigma_{f: x\to\mathbb{1}(x>0)}(\mathbf{V_2}) \quad :\{d\}\times (T\times T) & 6\\
    &\mathbf{V}_d =\mathbf{Merge} (\mathbf{V_1}, \mathbf{V_2})\quad :\{d\}\times (T\cup (T\times T)) & 7\\
    \end{align*}
    \vspace{-2em}
    \caption{Algebraic representation for task in Example~3.}
    \label{fig:bigram}
\end{figure}

Step 1 generates the term index matrix where each term is the unigram. The $\mathbf{Sum}_1$ operation in Step 2 generates the term vector where $\mathbf{V_1}(d, t)$ is the unigram $t$ in document $d$. Steps 3--6 get the term vector $\mathbf{V_2}$ where the column key set is all bigrams in $d$. Step 7 concatenates the two term vectors to get the representation for $d$.

For each document $d_i$ in collection $D = \{d_1, \cdots, d_n\}$, we get its  term vector $\mathbf{V_{di}} : \{d_i\}\times (T_i\cup (T_i\times T_i)) \to  R$ using the above steps, then apply the $\mathbf{Merge}$ operation to get the document-term matrix $\mathbf{M} : D\times T \to R$ where $T = (T_1\cup \cdots \cup T_n) \cup  ((T_1\times T_1)\cup \cdots \cup (T_n\times T_n))$ is the union of all unigrams and bigrams in the whole corpus,\\

    $\mathbf{Merge}(\mathbf{V_{d1}}, \mathbf{Merge}(\mathbf{V_{d2}}, \cdots, \mathbf{Merge}(\mathbf{V_{d(n-1)}}, \mathbf{V_{d(n)}}))).$\\
    
Besides word-occurrence as the values  of term document matrix, one can also use a term's tf-idf value. If all terms are considered, term document matrix $\mathbf{M}$ would be  of high dimension and sparse, which would be  costly to manipulate. A simple and commonly adopted method to reduce dimension is to select out informative words. The following presents the queries  to get   document-term  matrix $\mathbf{M}$ with the tf-idf values for only informative terms where the informativeness is measured by idf value. 

\noindent \textbf{Example 4.}
Given a collection of documents $D$, we have to generate a document-term matrix $\mathbf{M}$ for the top 1000 ``informative words'' where $\mathbf{M}(d, t)$ is the tf-idf value for term $t$ in document $d$. Suppose there is a term-document matrix $\mathbf{M_1}$ which stores the occurrence for all unigrams in each document (the construction is similar to that of  example 2 and thus is skipped),  $\mathbf{M}$ can be generated by the following steps. The function $idf$ in Step 3 is to calculate idf value, which is defined as $idf(x) = -\log\frac{x}{|D|}$ where $x$ is the number of documents that contains a specific term. 

\begin{figure}[ht]
    \centering
    \begin{align*}
        & \mathbf{M}_2 = \mathbf{Apply}_{f: x\to \mathbb{1}(x>0)} (\mathbf{M}_1)  \quad : D\times T& 1\\
        & \mathbf{V} = \mathbf{Sum}_1(\mathbf{M_2}) \quad : \{1\} \times T &2\\
        & \mathbf{I} = \sigma_{f:x\to \mathbb{1}(x\leq 1000)}(\mathbf{Sort}(\mathbf{V}))\quad : \{1\} \times T' & 3\\
        &\mathbf{V_1} = \mathbf{Apply}_{idf}(\Pi_{:, \mathbf{I}.K_2}(\mathbf{V}))\quad : \{1\} \times T' & 4 \\
        & \mathbf{M_3} = \Pi_{:, \mathbf{I}.K_2} (\mathbf{M_1})\quad :  D\times T' & 5\\
        & \mathbf{M} = \mathbf{Expand}_{\odot}(\mathbf{V_1}, \mathbf{M_3})\quad :  D\times T' & 6
    \end{align*}
    \caption{Algebraic representation for task in Example~4.}\label{fig:eg4}
\end{figure}

\subsection{Using TAAs}
For Example 1 introduced in Section~\ref{sec:intro}, we express this analysis using relational algebra and the associative array operations.  Suppose that the maximum number of words for  a term in $L_o \cup L_p$ is 3, now this analysis can be expressed as the following. The Step 1 is expressed in relational algebra. $\mathbf{TopicModel}$ in the last step is a function which takes a document-term matrix and produce  document topic matrix and topic term  matrix, which are the standard outputs of topic modeling, represented by another two TAAs $\mathbf{DTM}$ and  $\mathbf{TTM}$. Let $\mathbf{T} = \rho_{f:x\to \mathbb{1}(x\geq |D|-k)}(\mathbf{Sort}_2(\mathbf{DTM}))$, then $\mathbf{T}.K$ will return all $(d, t)$ pairs where $t$ is one of the top-$k$ topics for $d$.      

\begin{figure}[h]
    \centering
    \begin{align*}
    & D = \pi_{content} (\sigma_{d_1 \leq data \leq d_2} (R))  &1\\
    & \mathbf{M}: \{\}\times \{\}\to R,\quad  \mathbf{FV}: \{\}\times \{\}\to R &2\\
    & \text{for } d \in D: &3\\
    & \quad \quad \mathbf{N_1} = \mathbf{Tokenize}(d)  &3.1\\
        & \quad\quad \mathbf{V} = \rho_{\{1\} \to \{d\}, \{d\}\times T \to T}(\mathbf{Sum}_2(\mathbf{N_1}))^\mathsf{T} & 3.2\\
    &\quad \quad \mathbf{N_2} = \mathbf{Union}(\mathbf{N_1}, \mathbf{LShift_1(\mathbf{N_1})}) &3.3 \\
    & \quad\quad \mathbf{N_3} = \mathbf{Union}(\mathbf{N_2},\mathbf{LShift_2(\mathbf{N_1})} ) &3.4\\
    &\quad \quad \mathbf{N}= \mathbf{Merge}(\mathbf{N_1}, \mathbf{Merge}(\mathbf{N_2}, \mathbf{N_3})) &3.5\\
    &\quad \quad  \mathbf{V_f} = \rho_{\{1\} \to \{d\}, \{d\}\times T' \to T'}(\mathbf{Sum}_2(\mathbf{N})^\mathsf{T})& 3.6\\
    & \quad\quad \mathbf{FV} = \mathbf{Merge}(\mathbf{FV}, \mathbf{V_f}) &3.7\\
    & \quad \quad \mathbf{M} =  \mathbf{Merge}(\mathbf{M}, \mathbf{V}) &3.8\\
    & \mathbf{FV_o} =  \mathbf{\Pi}_{:, L_o}(\mathbf{FV}) &4\\
    & \mathbf{FV_p} = \mathbf{\Pi}_{:, L_p}(\mathbf{FV}) &5\\
    &\mathbf{I_o} = \sigma_{f:x\to\mathbb{1}(x>c_1)}(\mathbf{Sum}_2(FV_o)) &6\\
    &\mathbf{I_p} = \sigma_{f:x\to\mathbb{1}(x>c_2)}(\mathbf{Sum}_2(FV_p)) &7\\
    &\mathbf{M} = \Pi_{I_o.K_1\cap I_p.K_1,  :}(\mathbf{M}) &8\\
    &\mathbf{I_t} = \sigma_{f:x\to\mathbb{1}(x<\theta_2)}(\mathbf{Sum}_1(\mathbf{M}))&9\\
    &\mathbf{I_d}=\sigma_{f:x\to\mathbb{1}(x<\theta_1)}(\mathbf{Sum}_2(\mathbf{M}))&10\\
    &\mathbf{M} = \Pi_{I_d.K_1, I_t.K_2}(\mathbf{M}) &11\\
    &\mathbf{DTM}, \mathbf{TTM} = \mathbf{TopicModel(\mathbf{M})} & 12
\end{align*}
    \caption{Algebraic representation for the task in Example~1.}
    \label{fig:eg1}
\end{figure}
\vspace{-3mm}

% \begin{acks}
% \end{acks}

\bibliographystyle{ACM-Reference-Format}
\bibliography{sample-base}

% \appendix


\end{document}